# Coexistence of Superconductivity and Antiferromagnetism Probed by Simultaneous NMR and Electrical Transport in (TMTSF)$_2$PF$_6$

I.J. Lee[1,2], S.E. Brown[3], W. Yu[3], M.J. Naughton[4] and P.M. Chaikin[1]

[1]*Department of Physics, Princeton University, Princeton, NJ 08544, USA*
[2]*Department of Physics, Chonbuk National University, Jonju, 561-756, Korea\**
[3]*Department of Physics and Astronomy, UCLA, Los Angeles, CA 90095, USA*
[4]*Department of Physics, Boston College, Chestnut Hill, MA 02467, USA*

(Dated March 31, 2005)

We report simultaneous NMR and electrical transport experiments in the pressure range near the boundary of the antiferromagnetic spin density wave (SDW) insulator and the metallic/superconducting (SC) phase in (TMTSF)$_2$PF$_6$. Measurements indicate a tricritical point separating a line of second order SDW/metal transitions from a line of first order SDW/metal(SC) transitions with coexistence of macroscopic regions of SDW and metal(SC) order, with little mutual interaction but strong hysteretic effects. NMR results quantify the fraction of each phase.

The competition between superconductivity and magnetism has a long history. In it's most recent incarnation for many unconventional superconductors: high-temperature superconductors [1], Ce-based heavy fermions [2], ferromagnetic superconductors UGe$_2$ [Ref. 3] and ZrZn$_2$ [Ref. 4], and low dimensional organic systems [5,6], it is believed that the pairing interactions are of magnetic origin. The phase diagrams of these systems are quite similar, in the sense that the superconductivity emerges after the magnetic phase is suppressed or nearly suppressed by either the application of pressure or the doping of carriers. In all of the unconventional superconductors, it is of interest to understand the competition or collaboration of the magnetic and superconducting states. In the case of the organic superconductor (TMTSF)$_2$PF$_6$, an antiferromagnetic (spin density wave SDW) phase is directly adjacent to the triplet superconducting phase. Previous studies have suggested either second order transitions with a region of microscopic coexistence, or a first order transition and even a reentrant SDW phase in the region of the phase diagram where the two competing phases meet.

In a comprehensive paper, Vuletic *et al.* [7] suggest theoretically and from their experiments that a region of macroscopic coexistence of SDW and metal(SC) phases exists at temperatures below a tricritical point. More recently, Podolsky *et al.* [8] find similar coexistence within an SO(4) theoretical treatment, as do Zhang and Sa de Melo within a variational free energy approach [9]. Our results support this idea, showing explicitly from simultaneous NMR and transport near this tricritical point that the two phases coexist in the same sample [10], but in spatially separate regions, and we analyze the volume fractions of the two phases as a function of temperature. The observation that the angle-dependent magnetoresistance of the metallic phase as well as the critical temperatures of the two phases are unaffected by the existence of the competing phase shows that the domain sizes are characteristically larger than the mean free path in the metallic phase, and the correlation lengths in the ordered regions.

The low dimensional organic salt (TMTSF)$_2$PF$_6$ is renowned for its remarkably rich physical properties [11]. These properties range from an insulating to a superconducting phase, depending on applied pressure, magnetic field and temperature. Interest in the nature of the superconductivity has risen in the wake of recent reports of spin triplet superconductivity found from an upper critical field study [12] and an NMR Knight shift experiment [13]. The versatility of the organic system lies mainly in its highly anisotropic nature. Its quasi-one-dimensional Fermi surface consists of a pair of slightly warped sheets with bandwidths given by $4t_a$: $4t_b$: $4t_c$ = 1: 0.1: 0.003eV where $t_i$ are electron transfer energies along the *a*, *b*, and *c*-axes, respectively. Hereafter, we use *a*, *b* and *c* to represent the orthogonal *a*, *b'*, and *c$^*$* axes. In the (TMTSF)$_2$PF$_6$ system, superconductivity occurs near 1 K after the SDW insulating phase is suppressed by applying pressure above a critical pressure of ~6 kbar.

The phase regime in which we are interested is just above the critical pressure where Greene and Engler [14] previously suggested a possible formation of a mixed phase in the superconducting regime and called for more detailed study. Yamaji [15] considered the problem theoretically and concluded that the superconducting phase does not locally coexist with the SDW phase. In Yamaji's calculations, the uniform superconducting phase is completely separated from the SDW by a first-order phase transition. Recently, several groups have offered a new interpretation of the interesting phase regime, based on detailed electrical transport studies. Lee, *et al.* found a strong upward curvature in the critical field phase diagram [16,17], and were able to understand it with a simple model which involves self-consistently dividing superconductors into thinner layers in applied magnetic fields. In a study of temperature-dependent resistivity, as well as superconducting critical currents, Vuletic *et al.* [7] attempted to quantify the metallic volume fraction, under the assumption that the measured electrical resistivity is composed of two mutually independent sections of the SDW and the metallic phase. Quantification of the volume fractions using NMR linewidths as a local probe was presented in Ref. 10. Very recently, Kornilov, et al. [18] also studied mostly the c-axis magnetoresistance effect in the inhomogeneous (SDW/metallic) regime. In this report, we address the issue in a systematic manner by utilizing simultaneous proton NMR and electrical transport measurements integrated with results from an angular magnetoresistance oscillation (AMRO) study. AMRO directly probes the Fermi surface of the metallic phase, while NMR probes the properties of the SDW through the interaction of the nuclear spin with its local magnetic field.

A high quality $(TMTSF)_2PF_6$ single crystal, grown by standard electrocrystallization techniques, was mounted on the electrical feed-through of a BeCu pressure cell. The pressure cell was then loaded onto a sample holder – a string driven vertical rotator which was thermally anchored to the mixing chamber of a dilution refrigerator. In combination with a goniometer drive which rotated the entire dilution refrigerator, the vertical rotator provided $4\pi$ steradian rotations in a horizontal magnetic field. To perform simultaneous NMR and electrical transport measurements, a pressure cell with sufficiently large sample space (4.5 mm diameter, 4 mm length) was used, and a small sample wired for four-probe resistivity was placed inside an NMR coil. Our proton spin echo signal was obtained with $\pi/2$ - $\pi/3$ pulse sequences, presuming that the dominant spin-phase relaxation is caused by the dipolar interaction among rapidly tunneling methyl protons [19]. For $T>1K$, the rf pulse conditions were set such that the $\pi/2$ pulse duration was 2.2 μs. Lower power levels were used for $T<1K$.

In Fig. 1, we show simultaneous resistivity and proton NMR measurements under a pressure of 5.5 kbar and in a magnetic field of 0.29 T. Data with triangles (circles) were obtained with field along the a-axis (tilted 45 degrees toward the c-axis). The SDW transition, at a temperature near 3 K, was observed in all three types of measurements. The top panel shows the results from interlayer (c-axis) electrical transport, in which the resistance is enhanced below 3 K due to the SDW transition, followed by a superconducting transition near 1 K. At zero field (not shown), the resistance is increased by an order of magnitude, but was not thermally activated, suggesting the presence of a relatively large fraction of metallic phase before the superconducting phase was reached. As shown in the figure, the superconducting transition near 1 K was suppressed to ~0.5 K as the magnetic field was tilted toward the least conducting c-axis.

Our main focus was to obtain direct evidence for the presence of both SDW and metallic (SC) phases coexisting in the same sample. Therefore, we simultaneously measured the proton spin-lattice relaxation rate ($1/T_1$) (Fig. 1- middle panel). In the metallic state above 3K, a single exponential curve describes fairly well the recovery of the magnetization. Below 3K, the recovery deviates significantly from the form at higher temperatures as local variations of the spectral density develop. The diffusion of the nuclear spin magnetization makes a quantitative analysis of the recovery impractical, so we simply define $T_1$ using $M(T_1)=M_0(1-1/e)$, with $M_0$ the equilibrium value. Note that the superconducting transition is not evident here. The bottom panel shows the temperature dependence of the local magnetic field at the proton site, which is essentially the measure of the full width of the NMR absorption spectra shown in Fig. 2. The line width in the normal state, which is nearly independent of temperature and mostly due to nuclear dipolar coupling between methyl protons, was subtracted. The line broadening at low temperatures is solely associated with local field changes from the SDW state. The additional linewidth is proportional to the SDW order parameter.

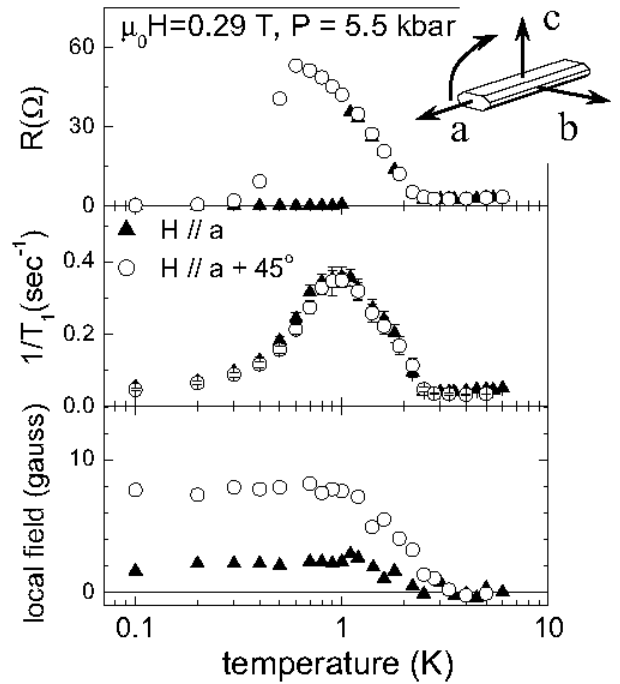

FIG. 1. Simultaneous resistivity and proton NMR measurements. Shown here are, from the top panel, the temperature dependence of interlayer resistance, proton spin-lattice relaxation rate and local field variations at the proton site. The data with triangles were obtained with a magnetic field aligned along the a-axis and circles with a 45 degree tilt toward the c-axis.

From Fig. 1, we note that the NMR results were dominated by the SDW signal. Moreover, the SDW, as seen by NMR, is largely unaffected by the superconducting state. This indicates that the SDW and superconducting regions are macroscopically separated (i.e. each domain is larger than its respective correlation length). NMR absorption spectra, normalized by compensating for the temperature effect, are shown in Fig. 2. The lineshape spreads as temperature decreases below 3 K. It is clear that a larger volume fraction of the nuclear spins in the initial metallic state was under the influence of larger static magnetic moment, as the SDW phase grew. The inset shows that, by the time the temperature reaches 1 K, about 30 percent of the spectral weight of the total absorption in the normal state is redistributed to the wings, which we associate with the portion of the sample in the SDW phase. Phase segregation such as this will occur in the vicinity of a first order phase transition whenever pressure is no longer a good control parameter [10]. While trivial pressure gradients also lead to apparent phase segregation, this seems unlikely because it cannot explain the observed variation of SDW volume fraction with temperature [10,16] and magnetic field [16].

Further insight into the nature of the domain structure of the interpenetrating phases was obtained from detailed magnetotransport measurements, recorded at $T = 0.15$K. With the magnetic field aligned along the c-axis, we detected a series of kink structures periodic in inverse field, which we associate with the field-induced SDW state [11], one of the well known properties of quasi-one-dimensional metals. Data show that once superconductivity is destroyed by a sufficiently high magnetic field, a portion of the sample which was not involved in the SDW transition behaves as a uniform metallic phase that usually exists at much higher pressure, away from the pure SDW phase. Moreover, the magnetoresistance displays an unusual hysteresis upon field cycling. Similarly, when the sample is warmed from the superconducting state, hysteretic behavior in $R(T)$ is observed in the "normal" (non-superconducting) state below the SDW transition ~3 K. These hysteretic behaviors strongly suggest a coexistence of SDW and metallic phases which change their volume fraction and/or the number of SDW domains with applied field and temperature. This finding is consistent with the results of the proton NMR measurement. The enhanced resistivity observed upon increasing $T$ or decreasing $B$ is likely due to an extra SDW volume fraction that has been pinned by disorder or impurities.

Further characterization of the magnetic-field driven metallic phase at low temperature can be obtained by AMRO studies, results of which are shown in Fig. 3. Here we see two types of AMRO effects typical of metallic systems with slightly warped quasi-one-dimensional Fermi surfaces (Q1D-FS), and understood in terms of classical Boltzmann transport theory. In Fig. 3(a), the a-c-resonance effect, arising from an averaging-out the interlayer (c-axis) carrier velocity over a Q1D-FS, is seen. This effect was initially observed in $(TMTSF)_2ClO_4$ [20] and later in $PF_6$ [21] at 8.3 kbar in a purely metallic regime, at a pressure significantly far from the SDW phase. As magnetic field is tilted away from the a-axis toward c (i.e. as θ is increased from 0), electrons sweep through the $k_b$ direction, but their orbits are limited along $k_c$, de-pending on the degree of tilt. When an electron path along $k_c$ fits within an integer number of Brillouin zones, the c-component velocity effectively averages to zero. This averaging effect manifests itself as a peak in the AMRO, as seen in Fig. 3 (a) near θ=±15°. Panel (b) shows the so-called "third angular effect" AMRO in the a-b plane. The local resistivity minimum seen at φ=±17.5 degrees is insensitive to magnetic field strength, suggesting it is due to a topological effect of a Q1D-FS [22]. In fact, it is due to effective electrons near inflection points of the Q1D-FS [23,24], at which an electron's large initial velocity decays little over time and thus enhances the conductivity. This condition is met when the magnetic field is oriented nearly parallel to the velocity vectors near the inflection points of the Q1D-FS.

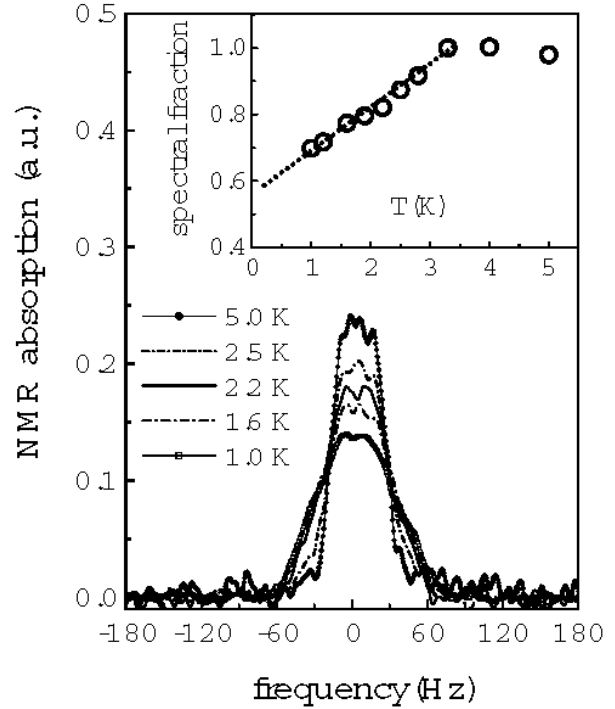

FIG. 2. Proton NMR line shapes at various temperatures and a pressure of 5.5 kbar. Each line is normalized by compensating the temperature effect. The inset indicates that the volume fraction of metallic spectral weight shifts to the periphery. The dotted line is guided by eyes. The nearly temperature independent volume fraction below 1 K is not shown.

From detailed calculations based on the Boltzmann transport equation, the transfer energy ratio $t_b/t_a$, a quantity proportional to the FS bandwidth, is found to be 1/8.9. The derived bandwidth is slightly reduced in comparison with the pure metallic case at 8.3 kbar [24]. These transport measurements, showing clear metal-phase AMRO effects, strongly suggest that the multiply connected superconducting/metallic and SDW domains are larger than several microns in dimension or at least

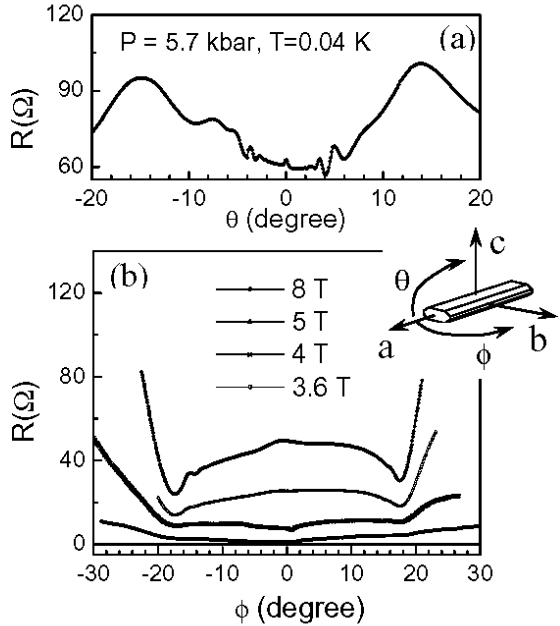

FIG. 3. Angular dependence of magnetoresistance obtained in the $a$-$c$ plane (panel (a)) and $a$-$b$ plane (panel (b)) at $P = 5.7$ kbar. The $a$-$c$-resonance in (a) was measured at 8 T.

compatible with the mean free path ($\sim \mu$m) of the uniform metallic phase.

The reason for the tricritical point in the $P$-$T$ phase diagram of $(TMTSF)_2PF_6$ is easily seen in terms of a Landau free energy expansion: $F_{SDW} - F_n = a(T - T_c)\phi^2 + u_4\phi^4 + u_6\phi^6$, where $F_{SDW}$ and $F_n$ are the free energy densities of the SDW and normal metallic states, respectively, $\phi$ is the SDW order parameter, $T_c$ the temperature at which the generalized susceptibility diverges, and $a$, $u_4$, and $u_6$ are system-dependent constants. Positive $u_4$ and $u_6$ are characteristic of a second order transition. For $u_4$ negative, the transition is first order. If $u_4$ changes sign as extrinsic parameters are changed, then $u_4=0$ marks the tricritical point passing from a line of second order to a line of first order transitions. This often occurs when a transition temperature depends on a quadratic degree of freedom ($x$), such as strain. Following the more detailed calculation of Vuletic et al. [7], we take $T_c(x) = T_{c0} - xT_c'$, $T_c' = \partial T_c / \partial x$ and add an elastic energy $\frac{1}{2}Kx^2$ to $F_{SDW}$. Minimizing with respect to $x$ we have: $x = -aT_c'\phi^2/K$ and $F_{SDW} - F_n = a(T-T_c)\phi^2 + (u_4 - a^2(T_c')^2/2K)\phi^4 + u_6\phi^6$. If $T_c' \to \infty$ as $T_c \to 0$ the coefficient of the quartic term goes through zero and we have a tricritical point. Generally, if $T_c$ varies with a control parameter as $T_c \sim |g-g_c|^{\nu z}$, and such quadratic coupling exists, then the transition is driven first order if $\nu z < 1$

In conclusion, using simultaneous independent probes of the SDW, superconducting and normal states in $(TMTSF)_2PF_6$, we confirm and quantify the coexistence of macroscopic segregated regions of SDW and metal below a tricritical point in $P$-$T$ space. There is little or no effect of any of one ordered phase on another, due to their spatial separation. We suggest that such tricritical behavior may be present in many other systems with parametric coupling to a quadratic degree of freedom.

This work was supported by the Korean Research Foundation Grant (KRF-2004-015-C00142) and research funds of Chonbuk National University (IJL) and by the National Science Foundation under Grant Numbers DMR-0243001 (PMC), DMR-0203806 (SEB), DMR-0308973 (MJN).